\documentclass[finals,3p]{elsarticle}

\usepackage{amssymb}
\usepackage{bm}

\journal{Physics Letters B}

\graphicspath{{figures/}}

\begin{document}

\begin{frontmatter}



\title{The nucleon properties in finite temperature and density with Gaussian fluctuations}


\author[first,second,third]{Peixin Weng}
\author[first,second,third]{Bingtao Li}
\author[first,second,third]{Yiming Lyu}
\author[fourth]{Song Shu}\ead{shus@hubu.edu.cn}
\author[first,third,fifth]{Hui Zhang}\ead{Mr.zhanghui@m.scnu.edu.cn}

\address[first]{State Key Laboratory of Nuclear Physics and Technology, Institute of Quantum Matter, South China Normal University, Guangzhou 510006, China}
\address[second]{Key Laboratory of Atomic and Subatomic Structure and Quantum Control (MOE), Guangdong-Hong Kong Joint Laboratory of Quantum Matter, Guangzhou 510006, China}
\address[third]{Guangdong Basic Research Center of Excellence for Structure and Fundamental Interactions of Matter, Guangdong Provincial Key Laboratory of Nuclear Science, Guangzhou 510006, China }
\address[fourth]{School of Physics, Hubei University, Wuhan, Hubei 430062, China}
\address[fifth]{Physics Department and Center for Exploration of Energy and Matter, Indiana University, 2401 N Milo B. Sampson Lane, Bloomington, IN 47408, USA}

\begin{abstract}
We investigate the properties of nucleons at finite temperature and density using a two-flavor quark meson model with Gaussian fluctuations that extend beyond the mean-field approximation.
Our findings suggest that Gaussian fluctuations lead to a non-monotonic behavior of the nucleon mass as a function of temperature and density, which may play an important role in the study of the hadronization process of relativistic heavy-ion collisions. Moreover, we observe an increase in the nucleon radius due to Gaussian fluctuations, suggesting an effective repulsive force akin to the Casimir effect, as observed in the gold-bromobenzene-silica system. This study offers new insights into how temperature, density, and quantum fluctuations affect the structure and properties of nucleons under extreme conditions.

\end{abstract}




\end{frontmatter}


\section{Introduction}

Quantum Chromodynamics (QCD) is the fundamental theory of strong interactions.  It is well established that QCD matter exhibits a hadron gas phase at low temperature and low density, transitioning to a quark-gluon plasma (QGP) phase at high temperature or density \cite{Braun-Munzinger:2008szb,Luo:2015doi}. While perturbative QCD is effective in high-energy or short-distance regimes, its non-perturbative nature, such as spontaneous chiral symmetry breaking and confinement, necessitate alternative approaches for low-energy or long-distance scenarios \cite{Collins:2011zzd,Ghiglieri:2020dpq,CTEQ:1993hwr,Deur:2016tte,Bethke:2006ac}. Lattice QCD has emerged as a vital non-perturbative tool for exploring the properties of hadronic matter \cite{LHPC:2007blg,Hagler:2009ni,Liang:2019frk}. However, because of the so-called sign problem, it encounters significant challenges in high-density regions. This limitation has led to the development of effective models that capture the essential features of QCD, including the Nambu-Jona-Lasinio (NJL) model~\cite{Klevansky:1992qe, Hatsuda:1994pi, Fukushima:2003fw, Fukushima:2008wg}, the MIT bag model~\cite{Chodos:1974je, Alford:1997zt, Johnson:1975zp}, the Hadron Resonance Gas (HRG) model~\cite{Hagedorn:1965st, Becattini:2000jw, Andronic:2005yp, Huovinen:2009yb}, the Chiral Effective Field Theory ($\chi$EFT)~\cite{Weinberg:1978kz, Gasser:1983yg, Ecker:1994gg, Scherer:2002tk}, and so on.

Experimental efforts, particularly through heavy-ion collisions at facilities such as the Relativistic Heavy Ion Collider (RHIC) at Brookhaven National Laboratory (BNL) and the Large Hadron Collider (LHC) at CERN, have significantly advanced our understanding of hadronic properties \cite{Hayano:2008vn,ALICE:2020mfd,Chen:2024aom,Shou:2024uga}. Furthermore, astronomical observations of compact stars help constrain the equation of state for hadronic matter \cite{Baym:2017whm}. The extreme environments created in heavy-ion collisions and within compact stars (characterized by high temperature, high density, rotation, strong magnetic fields, etc.) substantially alter the properties of hadrons. For instance, recent RHIC experiments have achieved collision energies in the $O(1) \ \rm{GeV}$ range \cite{STAR:2024znc}, resulting in considerable densities of nuclear matter. The density of nuclear matter in the center of neutron stars is also large. Under such high-density conditions, the wave functions of hadrons overlap, enhancing (both quantum and thermal) fluctuations and highlighting their critical role in determining hadronic properties.

The properties of hadrons have been extensively studied within the Quark Meson (QM) model over the past few decades. The thermodynamic properties of this model under various backgrounds have been investigated \cite{Petropoulos:1998gt, Nemoto:1999qf, Scavenius:2000qd, Baacke:2002pi, Mocsy:2004ab, Bowman:2008kc, Zhang:2015vva, Zhang:2017icm, Osman:2024xkm}. Its extension with the Polyakov loop, accounting for the contribution of the gluon field and acting as an order parameter for confinement, has also been thoroughly examined~\cite{Schaefer:2007pw, Skokov:2010wb, Skokov:2010uh, Herbst:2010rf, Herbst:2013ail, Fukushima:2017csk}. The mass and radius of nucleon have garnered significant attention in both theoretical and experimental research\cite{Tanihata:1985psr,Abrahamyan:2012gp,Gallimore:2024fcz}.The QM model possesses a semiclassical soliton solution that can be considered as a chiral soliton ~\cite{Birse:1984loi, Birse:1984js} (a bound state of valence quarks), providing a successful approach for describing the static properties of nucleons~\cite{Alberto:1988xj, Bernard:1988db, Aly:1998wg}. Recent studies have demonstrated its success in describing meson and nucleon properties in both vacuum and thermal medium~\cite{Naar:1993cy, Abu-Shady:2012ewe, Mao:2013qu, Zhang:2014wta, Abu-Shady:2015ava, Jin:2015goa, Li:2018rfu, Wang:2023omt}. Previous investigations primarily relied on the mean-field approximation, which neglects significant quantum fluctuations. However, higher-order contributions may play a crucial role. 

In light of this, we examine the impact of fluctuations on the properties of nucleons beyond the mean-field approximation, which has previously been overlooked. Our results demonstrate that both the nucleon mass and radius exhibit significant changes near the boundary of the chiral phase transition, providing new insights into the behavior of hadronic matter under extreme conditions. The structure of this paper is as follows. In Sec.~\ref{sec:model}, we introduce the quark-meson model with two-flavor quarks and calculate the thermodynamic potential in the mean field, incorporating Gaussian fluctuations. In Sec.~\ref{sec:soliton}, we derive the soliton field equations in spherical coordinates and determine the static nucleon properties through the soliton solution. In Sec.~\ref{sec:result}, we present the numerical results and discussion. Finally, we provide a brief summary in Sec.~\ref{sec:summary}.

\section{Model}
\label{sec:model}

The Lagrangian of the two-flavor quark meson model in Minkowski space-time is given by
\begin{equation}
	\mathcal{L}=\bar \psi [i \gamma^\mu \partial_\mu +g(\hat{\sigma} +i \gamma_5 \vec{\tau} \cdot \hat{ \vec {\pi}} )  ] \psi + \mathcal{L}_{Km} - U( \hat{\sigma} ,\hat{ \vec {\pi}}), \label{Lagrangian}
\end{equation}
where the meson kinetic energy is
\begin{equation}
    \mathcal{L}_{Km}=\frac{1}{2}(\partial_\mu \hat{\sigma} \partial^\mu \hat{\sigma} + \partial_\mu  \hat{ \vec {\pi}} \partial^\mu \hat{ \vec {\pi}}),
\end{equation}
and the potential at zero temperature and density is
\begin{equation}
	U(\hat{ \sigma } ,\hat{ \vec {\pi}})=\frac{\lambda}{4} ( \hat{\sigma}^2 + \hat{\vec{\pi}}^2 -\zeta^2 )^2+ H \hat{ \sigma }. \label{mesonpotential} 
\end{equation}
When the $H \hat{\sigma}$ term is nonzero, the chiral symmetry of the Lagrangian is explicitly broken. The scalar meson field $\hat{\sigma}$ acquires a nonzero vacuum expectation value $\sigma_v$, whereas the pseudo-scalar meson field $\hat{\vec{\pi}}$ maintains a zero vacuum expectation value. The nonzero vacuum expectation value $\sigma_v$ is determined by the following equation:
\begin{equation}
	\lambda \sigma_v(\sigma_v^2 -\zeta^2)-H=0.
\end{equation}
In Eq.~(\ref{Lagrangian}), there is no explicit current mass term for the quark field and quarks get their constituent mass $g \sigma_v$ through the vacuum explicitly broken. The parameters of this model are given by the following equations,
\begin{eqnarray}
	&&H= f_\pi m_\pi^2,  \qquad  \lambda=(m_\sigma^2-m_\pi^2)/2f_\pi^2, \nonumber \\
	&&\zeta^2 = f_\pi^2-m_\pi^2/\lambda,  \qquad  g=m_q/f_\pi,
\end{eqnarray}
where $f_\pi$ is pion decay constant $f_\pi=93 \; \rm{MeV} $ and $m_\pi=138 \; \rm{MeV} $ is the pion mass. In our calculation, we follow the Ref.~\cite{Birse:1984loi, Birse:1984js}, and set the quark mass and sigma mass as $m_q=500 \; \rm{MeV}$  and $m_\sigma=1200 \; \rm{MeV}$ respectively, which means $g \approx 5.28$ and $\lambda \approx82.1$. The parameters are used to reproduce the properties of nucleons, including the ``experimental" mass of 1086 MeV (an equal mixture of $\Delta$ and $N$), values of the magnetic moments $(2M/e)\mu_p$ and $(2M/e)\mu_n$.

\subsection{Thermodynamic Potential}

Using finite-temperature field theory methods, the partition function is expressed as
\begin{eqnarray}
	&\mathcal{Z}&=\mathrm{Tr} \mathrm{exp} \left\lbrace -\beta(\hat{\mathcal{H}}-\mu \hat{\mathcal{N}}) \right\rbrace  \nonumber \\ 
	& &= \int \mathcal{D} \bar{\psi} \mathcal{D}\psi \mathcal{D}\sigma \mathcal{D}\vec{\pi} \mathrm{exp} \left\lbrace \int_{0}^{\beta} d\tau \int_{V} d^3x  (\mathcal{L} +\mu \bar{\psi} \gamma^0 \psi) \right\rbrace.
\end{eqnarray}


In the mean-field approximation, the meson fields are replaced with their vacuum expectation values:
\begin{eqnarray}
	&\mathcal{Z}_{MF}&= \int \mathcal{D} \bar{\psi}\mathcal{D}\psi \mathrm{exp}\left\lbrace \int_{0}^{\beta} d\tau \int_{V} d^3x  (\mathcal{L}+\mu \bar{\psi} \gamma^0 \psi)\right\rbrace \nonumber \\
	& &=\mathrm{exp}(- \frac{VU}{T}) \mathrm{det_p} \left\lbrace [p_\mu \gamma^\mu +\mu \gamma^0 - g(\sigma + i\gamma_5  \vec{\tau} \cdot \vec{\pi} )]/T \right\rbrace .
\end{eqnarray}
The thermodynamic quantities can be derived from the partition function, leading to the thermodynamic potential:
\begin{equation}
	\Omega_{MF}=\Omega_{\bar{q}q}+U(\sigma, \vec{\pi}),
\end{equation}
where the contribution of (anti)quarks at finite temperature and density is given by
\begin{eqnarray}
	\Omega_{\bar{q}q} = \frac{-\nu_q}{2 \pi^2} \int_{0}^{\infty} dp p^2 \left\lbrace   E +T\ln[1+ \mathrm{exp}(\frac{\mu-E}{T})] +T\ln[1+\mathrm{exp}(\frac{-\mu-E}{T})] \right \rbrace \label{Omegaq}.
\end{eqnarray}
Here, $\nu_q=2 N_c N_f=12$ is the number of degrees of freedom for the quark fields and $E=\sqrt{p^2+M^2}$ is the energy of the constituent quark. The constituent quark mass is defined as 
\begin{equation}
	M=\sqrt{g^2(\sigma^2+\vec{\pi}^2)}.
\end{equation} 
In Eq.~(\ref{Omegaq}), the zero-point energy term $\int_{0}^{\infty} dp \ p^2 E$ is divergent and requires renormalization. We neglect this term in our calculations and focus on the contribution from the Gaussian fluctuations.

We find that the system is primarily determined by constituent quarks, while the contribution from scalar fields is always zero. Physically, the degrees of freedom associated with mesons is also significant. In this context, we follow Ref.~\cite{Mocsy:2004ab} to incorporate the fluctuations from meson fields.

\subsection{Gaussian Fluctuations}

It is convenient to decompose the scalar field and pseudo-scalar field into vacuum expectation value plus fluctuations: $\hat{\sigma} = \sigma_v + \Delta, \hat{\vec{\pi}}= \vec{\pi}_v + \vec{\delta}$. Here, $\Delta$ represents the fluctuation of $\hat{\sigma}$, and $\vec{\delta}$ represents the fluctuation of $\hat{\vec{\pi}}$, respectively. In this context, the average of any odd-order fluctuations is zero, indicating that
\begin{eqnarray}
	\langle \Delta^{odd} \rangle =0,\; \langle \Delta^{even} \rangle \neq 0,  \nonumber \\ 
	\langle \vec{\delta}^{odd} \rangle =0,\; \langle \vec{\delta}^{even} \rangle \neq 0. 
\end{eqnarray}
We use angle brackets to denote averaging over fluctuations, and the specifics of this averaging process will be discussed in the appendix.

As mentioned above, odd-order fluctuations cancel out on average. Consequently, only even-order fluctuations contribute to the thermodynamic potential, while odd-order fluctuations are associated with the dynamic part. The meson masses are defined as the second derivative of the thermodynamic potential with respect to the fields,
\begin{eqnarray}  
	m_\sigma^2 = \frac{\partial^2 \langle  \Omega(\sigma , \vec{\pi} ) \rangle}{\partial \sigma^2 },  \nonumber \\ 
	m_\pi^2 = \frac{\partial^2 \langle  \Omega(\sigma , \vec{\pi} ) \rangle }{\partial \pi^2 }.   \label{mesonmass}
\end{eqnarray}

We linearize the effective mesonic potential by performing the following operation:
\begin{equation}
	U(\sigma, \vec{\pi}) \rightarrow \langle  U(\sigma, \vec{\pi} ) \rangle + \frac{1}{2} m_\sigma^2( \Delta^2 - \langle \Delta^2 \rangle ) + \frac{1}{2} m_\pi^2( \vec{\delta}^2 - \langle \vec{\delta}^2 \rangle ).  \label{linearize}
\end{equation}
The terms on the right-hand side that have been averaged will be incorporated into the potential term, while the remaining terms ($\Delta^2$ and $\vec{\delta}^2$), along with the kinetic term of the meson field, contribute to the partition function of meson fields:
\begin{eqnarray}
	\mathcal{Z}_{m}=\int \mathcal{D} \Delta  \mathcal{D} \vec{\delta}  \mathrm{exp} \left\lbrace \int_{0}^{\beta} d\tau \int_{V} d^3x [\mathcal{L}_{Km}- \frac{1}{2}m^2_\sigma \Delta^2 - \frac{1}{2}m^2_\pi \vec{\delta}^2]\right\rbrace. \label{Omegam}
\end{eqnarray}
Here we need to distinguish each term in Eq.~(\ref{linearize}). The term $\langle U(\sigma,\vec{\pi}) \rangle$ represents the meson potential $U$, as defined in Eq.~(\ref{mesonpotential}). The term $\left\langle \Delta^2 \right\rangle$ and $\langle \vec{\delta}^2 \rangle$  can be regraded as the potential of quasi-particles. 

Using finite-temperature field theory methods, we can obtain the thermodynamic potential of mesons:
\begin{eqnarray}
	\Omega_m &=& - \frac{\ln Z_m }{\beta V} \equiv \Omega_\sigma + \Omega_\pi, \\
	\Omega_\sigma &=& \frac{T}{2\pi^2} \int dp p^2 \left\lbrace \frac{1}{2}  E_\sigma +\ln [1-\mathrm{exp}(- E_\sigma/T)] \right\rbrace ,\\ 
	\Omega_\pi &=& \frac{3T}{2\pi^2} \int dp p^2 \left\lbrace \frac{1}{2}  E_\pi +\ln [1-\mathrm{exp}(- E_\pi/T)] \right\rbrace,
\end{eqnarray}
where $E_\sigma=\sqrt{p^2+m_\sigma^2},\ E_\pi=\sqrt{p^2+m_\pi^2}$. Two consistency relationships between the meson masses and fluctuations are obtained from Eq.~(\ref{Omegam}),
\begin{eqnarray}
	\langle \Delta^2 \rangle = 2 \frac{\partial \Omega_\sigma }{\partial m_\sigma^2 }=\frac{1}{2\pi^2}\int dp \frac{p^2}{E_\sigma} \frac{1}{\mathrm{exp} \left[ E_\sigma/T \right]-1},  \nonumber\\
	\langle \vec{\delta}^2 \rangle = 2 \frac{\partial \Omega_\pi }{\partial m_\pi^2 }=\frac{3}{2\pi^2}\int dp \frac{p^2}{E_\pi} \frac{1}{\mathrm{exp} \left[ E_\pi /T \right]-1} .  \label{conden}
\end{eqnarray}

The total thermodynamic potential is given by:
\begin{eqnarray}
	\left\langle \Omega(\sigma,\vec{\pi}) \right\rangle =\left\langle \Omega_{\bar{q}q}(\sigma,\vec{\pi}) \right\rangle  +\left\langle U(\sigma,\vec{\pi}) \right\rangle   - \frac{1}{2} m_\pi^2 \langle \vec{\delta}^2\rangle  -\frac{1}{2} m_\sigma^2 \langle \Delta^2 \rangle + \Omega_m .
\end{eqnarray}

\subsection{Soliton Field Equation And Static Baryon Properties}  \label{sec:soliton}

From the Lagrangian in Eq.~(\ref{Lagrangian}), we derive the radial soliton field equations in spherical coordinates:
\begin{eqnarray}
	&& \frac{dv(r)}{dr}=-(\frac{2}{r} -g\pi(r))v(r)+(g\sigma(r)+\epsilon)u(r), \\
	&&\frac{du(r)}{dr}=-(\epsilon - g\sigma(r) ) v(r)-g\pi(r) u(r), \\
	&&Ng (u^2(r)-v^2(r)) + \frac{d^2 \sigma(r)}{dr^2}+ \frac{2}{r}\frac{d \sigma(r)}{dr}=\frac{\partial U( \sigma(r)  ,\pi(r))}{\partial \sigma},	\\
	&&Ng (2 u(r)v(r) )+\frac{d^2 \pi(r)}{d r^2} +\frac{2}{r} \frac{d \pi(r)}{dr} - \frac{2 \pi(r)}{r^2}=\frac{\partial U(\sigma(r) ,  \pi(r) )}{\partial \pi} .
\end{eqnarray}
In these equations, we adopt the ``hedgehog" ansatz, which implies:
%
\begin{eqnarray}
	&& \langle \hat{\sigma}(\vec{r},t) \rangle  =\sigma(r),\quad    \langle \hat{\vec{\pi}}(\vec{r},t) \rangle= \hat{\vec{r}}\pi(r), \\
	&& \psi(\vec{r},t)=e^{-i \epsilon t} \sum_{i=1}^{N} q_i (\vec{r}),\quad q(\vec{r}) =\left( \begin{array}{c}
		u(r) \\ i \vec{\sigma} \cdot {\hat{\vec{r}}} v(r)
	\end{array}  \right) \chi, \\
	&& (\vec{\sigma} +\vec{\tau})\chi=0. 
\end{eqnarray}	
Here, $\chi$ denotes a spinor, and $q_i$ represents $N$ identical valence quarks with the lowest eigenenergy $\epsilon$. This corresponds to a baryon composed of three identical valence quarks when $N =3$. The quark wave function must satisfy normalization conditions, with each quark contributing one-third to the baryon number:
\begin{eqnarray}
	4\pi \int r^2(u^2(r) +v^2(r)) dr=1 .
\end{eqnarray}
The boundary conditions are as follows:
\begin{eqnarray}
	&& v(0)=0,  \qquad  \left.\frac{d \sigma(r)}{dr}\right|_{r=0}=0,  \qquad  \pi(0)=0, \\
	&&  u(\infty)=0,  \qquad  \sigma(\infty)=\sigma_v,  \qquad  \pi(\infty)=0,
\end{eqnarray}
where $\sigma_v$ is the expectation value of  {$\hat{\sigma}$} field in zero temperature and density vacuum.  

To investigate the chiral soliton in a thermal environment, we embed a soliton into a hot and dense homogeneous background, described by a thermal effective potential, and introduce meson fluctuations based on the mean-field approximation. The effective Lagrangian is given by:
\begin{equation}
	\mathcal{L}_{eff}=\bar \psi [i \gamma^\mu \partial_\mu +g(\hat{\sigma} +i \gamma_5 \vec{\tau} \cdot \hat{ \vec {\pi}} )  ] \psi + \frac{1}{2}(\partial_\mu \hat{\sigma} \partial^\mu \hat{\sigma} + \partial_\mu  \hat{ \vec {\pi}} \partial^\mu \hat{ \vec {\pi}}) - \Omega(\hat{\sigma} , \hat{\vec {\pi}},T,\mu ), \label{effectiveLagrangian}
\end{equation}
where $\Omega(\hat{\sigma} , \hat{\vec {\pi}},T,\mu)$ is the thermodynamic potential under the given conditions. In our calculation, $\Omega(\hat{\sigma},\hat{\vec{\pi}},T,\mu)$ will be replaced by $\Omega_{MF}(\hat{\sigma},\hat{\vec{\pi}},T,\mu)$ if we use the mean-field approximation, and by $\langle \Omega(\hat{\sigma},\hat{\vec{\pi}},T,\mu) \rangle $ if we consider Gaussian fluctuations based on the mean field. 

The radial soliton field equations in spherical coordinates within a thermal background, including Gaussian fluctuations, are:
\begin{eqnarray}
	&&\frac{dv(r)}{dr}=-(\frac{2}{r} -g\pi(r))v(r)+(g\sigma(r)+\epsilon)u(r), \\
	&&\frac{du(r)}{dr}=-(\epsilon - g\sigma(r) ) v(r)-g\pi(r) u(r), 		\\
	&&Ng (u^2(r)-v^2(r)) + \frac{d^2 \sigma(r)}{dr^2}+ \frac{2}{r}\frac{d \sigma(r)}{dr}= \frac{\partial \left\langle \Omega(\sigma(r), \pi(r), T, \mu) \right\rangle }{\partial \sigma },  	\\
	&&Ng (2 u(r)v(r) )+\frac{d^2 \pi(r)}{d r^2} +\frac{2}{r} \frac{d \pi(r)}{dr} - \frac{2 \pi(r)}{r^2}=\frac{\partial \left\langle \Omega(\sigma(r), \pi(r), T, \mu) \right\rangle }{\partial \pi },  
\end{eqnarray}
where $\langle \Omega(\sigma, \vec{\pi}, T, \mu) \rangle $ is the thermodynamic potential after averaging over the fluctuations. In this case, the previous setting and boundary conditions do not need to be modified, except that $\sigma_v$ is replaced by the expectation value of the thermal medium.

We employ numerical methods to solve the soliton equations. Physical quantities of the triple quark system can be obtained from the soliton solutions. Starting with the Lagrangian, we can obtain the Hamiltonian of the system through a Legendre transformation:
\begin{equation}
	H=\bar \psi \left[ -i \bm{\gamma}\cdot \bm{\nabla} - g(\sigma +i \gamma_5 \vec{\tau} \cdot  \vec {\pi} )  \right]   \psi + \frac{1}{2}(|\nabla \sigma|^2+|\bm{\nabla} \vec{\pi}|^2) + \langle \Omega( \sigma, \pi, T,\mu )\rangle .
\end{equation}
For a given eigenstate of the Hamiltonian, the energy is
\begin{equation}
	E=\left\langle H \right\rangle =N \epsilon + 4 \pi \int dr r^2 [ \frac{1}{2}  (\frac{d\sigma(r)}{dr})^2 + \frac{1}{2}(\frac{d\pi(r)}{dr})^2 + \frac{\pi^2(r)}{r^2}  +\left\langle \Omega(\sigma(r), \pi(r),T,\mu)\right\rangle ].
\end{equation}
From this equation, it is evident that the thermal potential at finite temperatures and densities is nonzero. In order to maintain the convergence of the integral, it's necessary to shift the minimum of the potential to zero by adding a subtraction factor $B(T,\mu)$~\cite{Zhang:2015vva}, where $B(T,\mu)$ represents the value of the thermal vacuum at a given temperature and chemical potential. This modification implies that the pressure at the surface of the soliton remains zero in both the vacuum and the thermal medium. Consequently, after including Gaussian fluctuations, the energy of the system, which is also defined as the nucleon mass, is expressed as:
\begin{eqnarray}
	E = M_B = N \epsilon + 4 \pi \int dr r^2 [ \frac{1}{2}  (\frac{d\sigma(r)}{dr})^2 + \frac{1}{2}(\frac{d\pi(r)}{dr})^2 + \frac{\pi^2(r)}{r^2}  + \left\langle \Omega(\sigma(r), \pi(r), T, \mu)\right\rangle +B(T,\mu)  ].  
\end{eqnarray}
Additionally, the root mean square (RMS) radius is defined as
\begin{equation}
	R=\sqrt{\left\langle r^2 \right\rangle }=\sqrt{4\pi \int dr r^4(u^2(r)+v^2(r))} .
\end{equation}

\section{Numerical Result and Discussion}  \label{sec:result}

\begin{figure}[htb]
	\centering
	\includegraphics[width=.45\textwidth]{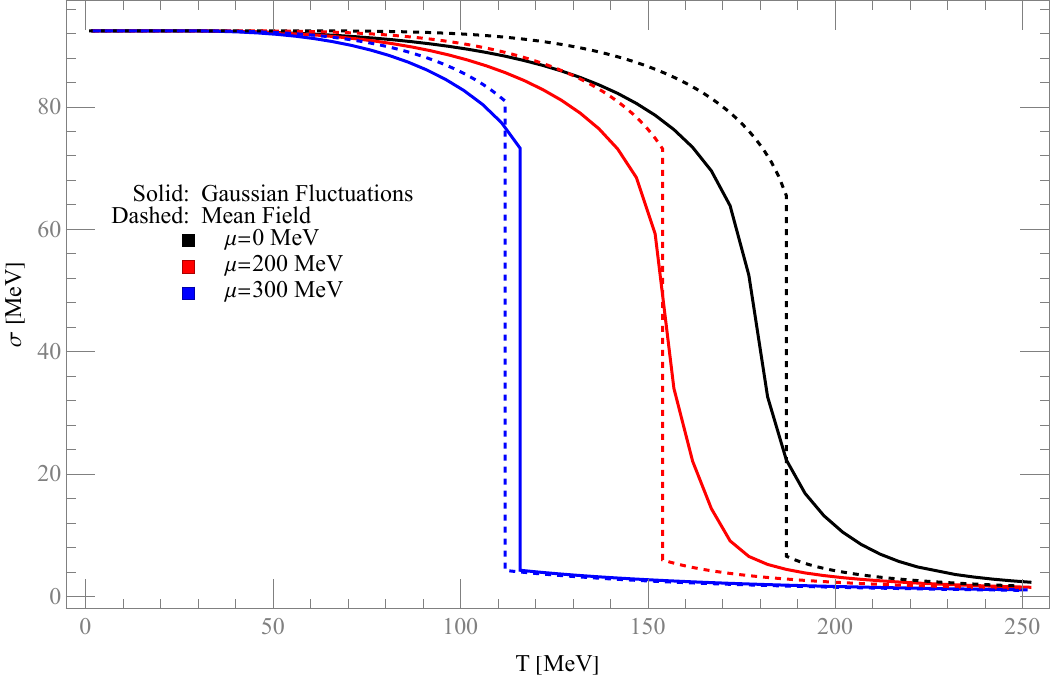}
	\includegraphics[width=.45\textwidth]{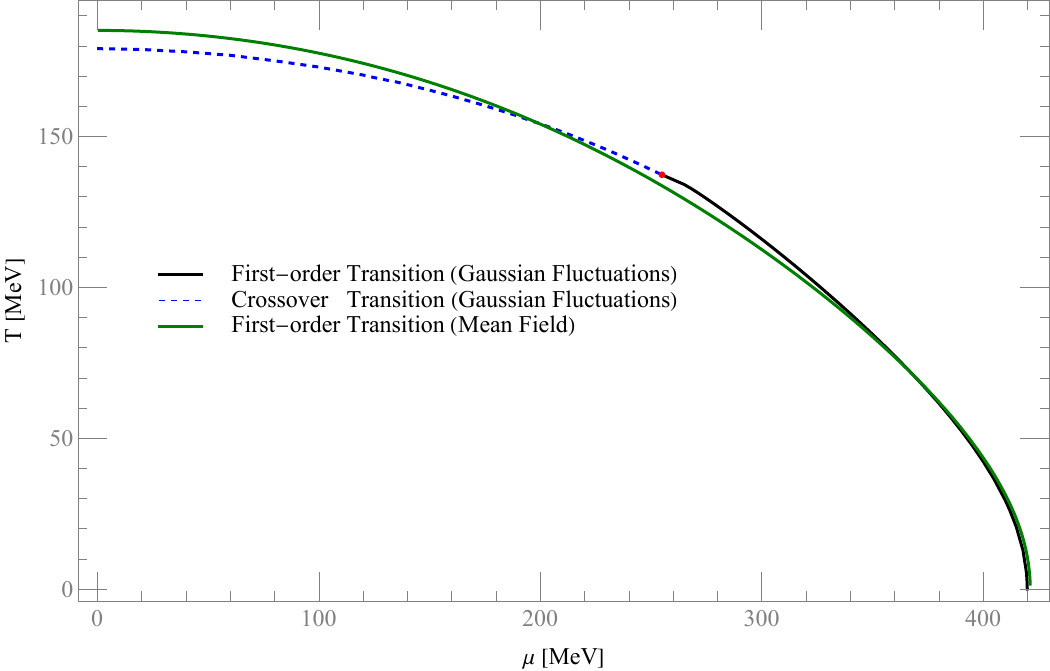}
	\caption{(left) Order parameter $\sigma$ as a function of $T$ with (solid lines) and without (dashed lines) Gaussian fluctuations, for $\mu= 0,\ 200,\ 300\ \rm{MeV}$. (right) The phase diagram in the $T-\mu$ plane for two-flavor quark meson model under two different approximations.  Solid (dashed) lines show the first order (crossover) phase transitions, while the dot marks the critical end point (CEP). } \label{phasediagram}
\end{figure}

The meson mass Eqs.~(\ref{mesonmass}), and the fluctuation Eqs.~(\ref{conden}), as well as the gap equation for the expectation value of the sigma field are coupled with each other, which requires us to solve them simultaneously. The thermodynamic potential can also be derived after averaging over fluctuations, which varies with meson masses $m_\sigma, m_\pi$, the chiral condensate $\sigma$,  {and the meson fluctuations $\langle \Delta^2 \rangle $, $\langle \delta^2 \rangle$}. In order to determine the location of the phase transition boundary, we compute the thermodynamic potential as a function of temperature $T$ and chemical potential $\mu$. In high temperature and low density regions, the three equations yield a single solution for a given $T$ and $\mu$. We identify the point of most rapid change in the order parameter as the crossover transition point. In contrast, at low temperature and high density regions, the three equations have non-unique solutions near the phase transition area. The thermodynamic potential forms two minima: one corresponding to the true vacuum state, and the other to a metastable state. When the metastable state transitions to the true vacuum, a first-order phase transition occurs. The left panel of Fig.~\ref{phasediagram} illustrates the order parameter as a function of temperature and density, highlighting this feature.

In the right panel of Fig.~\ref{phasediagram}, we compare the phase diagram of the quark meson model with the mean-field method and with Gaussian fluctuations in the $T-\mu$ plane. The system exhibits both first-order and crossover transitions due to the presence of fluctuations, rather than only first-order transitions as predicted by the mean-field approximation. The location of the phase boundary with Gaussian fluctuations can change the order of phase transition. Similar behavior can be found in Ref.~\cite{Skokov:2010sf}, which incorporates the contribution of fermion vacuum fluctuations. The location of the Critical End Point (CEP) is found to be around $T_c = 137 \ \rm{MeV}$, $\mu_c = 255 \ \rm{MeV}$.

\begin{figure}[htb]
	\centering
	\includegraphics[width=.5\textwidth]{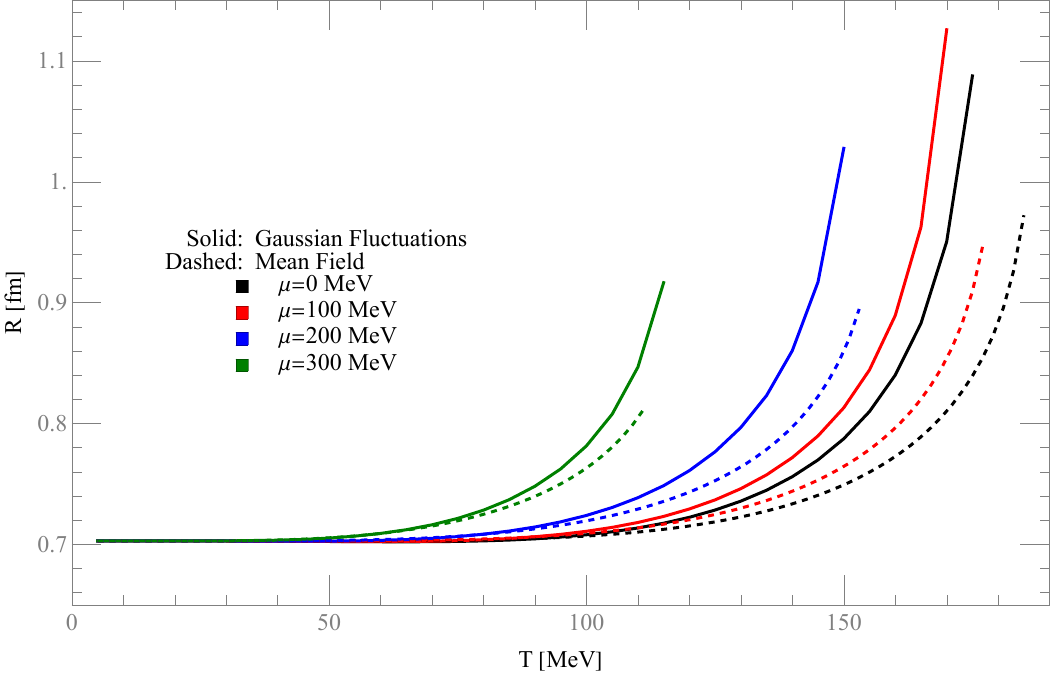}
	\caption{The RMS radius $R$ as a function of $T$ with (solid lines) and without (dashed lines) Gaussian fluctuations, for $\mu= 0,\ 100,\ 200,\ 300\ \rm{MeV}$. } \label{RT}
\end{figure}

In Fig.~\ref{RT}, we illustrate the RMS radius $R$ as a function of $T$ for different chemical potentials, $\mu=0,\ 100,\ 200,\ 300\ \rm{MeV}$. The radius, when accounting for Gaussian fluctuations, increases more rapidly than in the mean-field one, indicating that Gaussian fluctuations behave as a repulsive force. As we know, vacuum quantum fluctuations can produce a Casimir force in the gold-bromobenzene-silica system, essentially a repulsive one~\cite{Munday:2009fgb}. In our model, the addition of Gaussian fluctuations may provide a repulsive force similar to the Casimir effect, causing a more rapid rise in the radius with temperature.

\begin{figure}[htb]
	\centering
	\includegraphics[width=.5\textwidth]{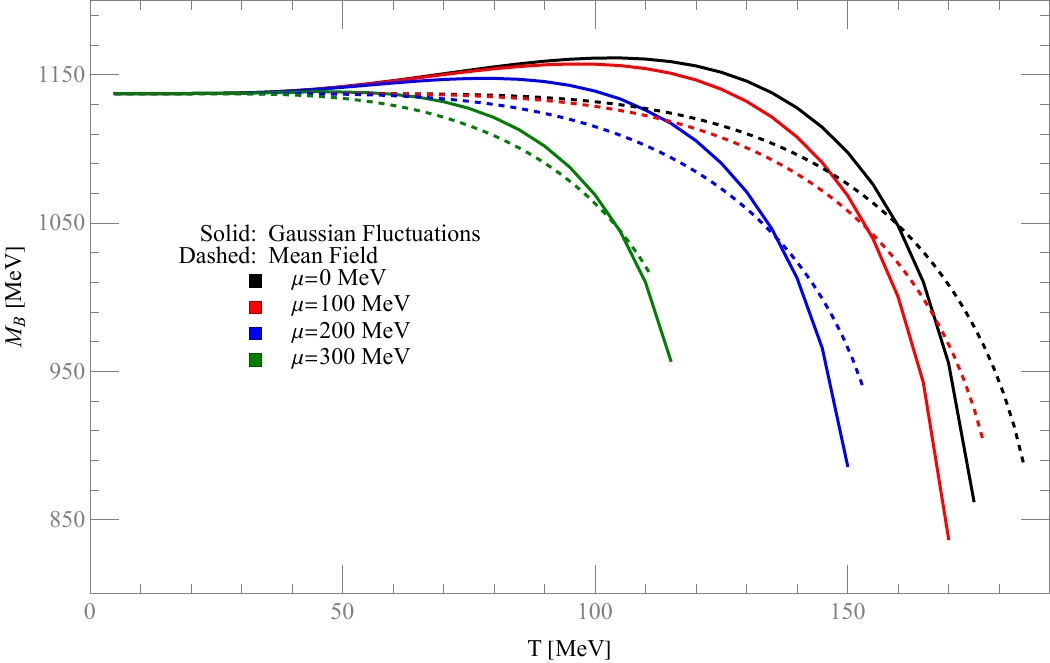}
	\caption{The nucleon mass $M_B$ as a function of $T$ with (solid lines) and without (dashed lines) Gaussian fluctuations, for $\mu = 0,\ 100,\ 200,\ 300\ \rm{MeV}$. } \label{MBT}
\end{figure}

In Fig.~\ref{MBT}, we plot the nucleon mass $M_B$ as a function of $T$ for different chemical potentials, $\mu= 0,\ 100,\ 200,\ 300\ \rm{MeV}$. Gaussian fluctuations cause $M_B$ to decrease more rapidly with temperature compared to the mean-field results. Moreover, $M_B$ exhibits an increase within the intermediate temperature range.
The significant change in hadronic mass near the phase boundary may play an important role in heavy-ion collisions. In previous studies, the mass of hadrons is typically fixed during the hadronization process in heavy-ion collision simulation, which is somewhat unphysical. The impact of variations in hadron mass on physical observables, such as the yield and its distribution, needs to be re-evaluated.

\begin{figure}[htb]
	\centering
	\includegraphics[width=.45\textwidth]{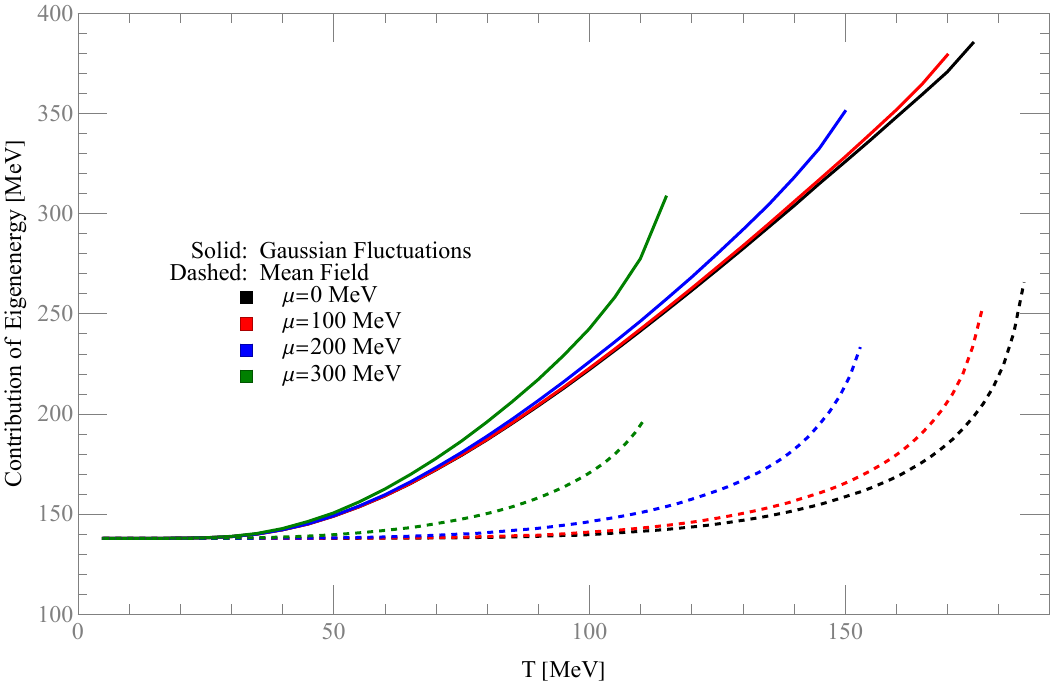}
	\includegraphics[width=.45\textwidth]{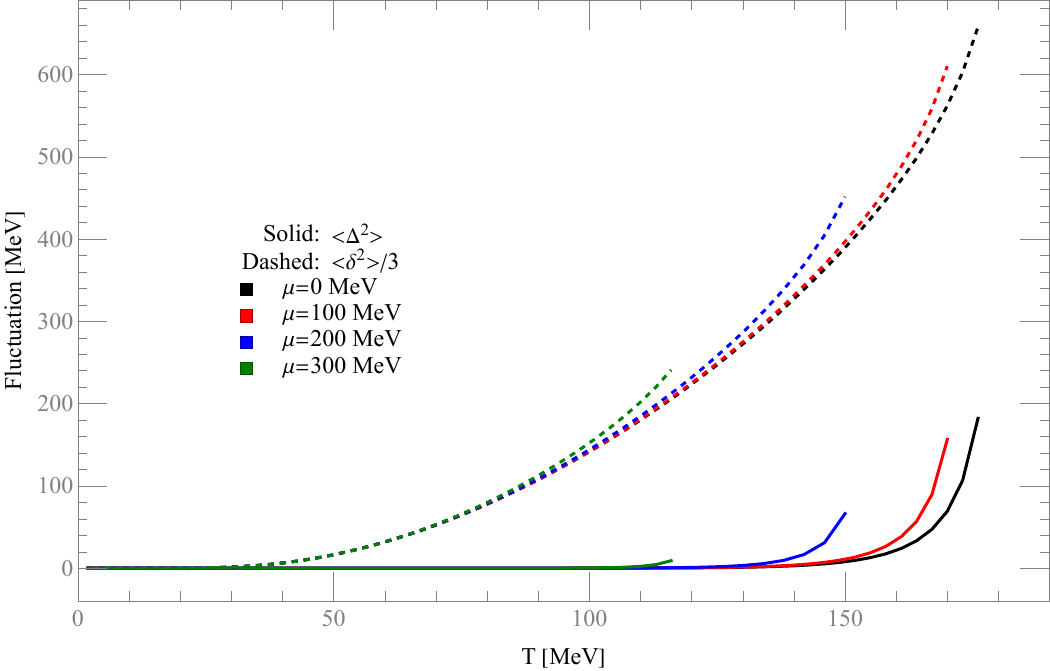}
	\caption{Eigenengergy (left) and fluctuations (right) as a function of $T$, for $\mu= 0,\ 100,\ 200,\ 300\ \rm{MeV}$. } \label{contributionall}
\end{figure}

In order to understand the non-monotonic behavior of $M_B$ in the intermediate temperature range, we decompose $M_B$ into contributions from quark field eigenenergy and meson fields. We find that the increase of $M_B$ is driven by the rapid rise in quark field eigenenergy, which may be strongly correlated with meson fluctuations, as observed in Fig.~\ref{contributionall}.

\begin{figure}[htb]
	\centering
	\includegraphics[width=.5\textwidth]{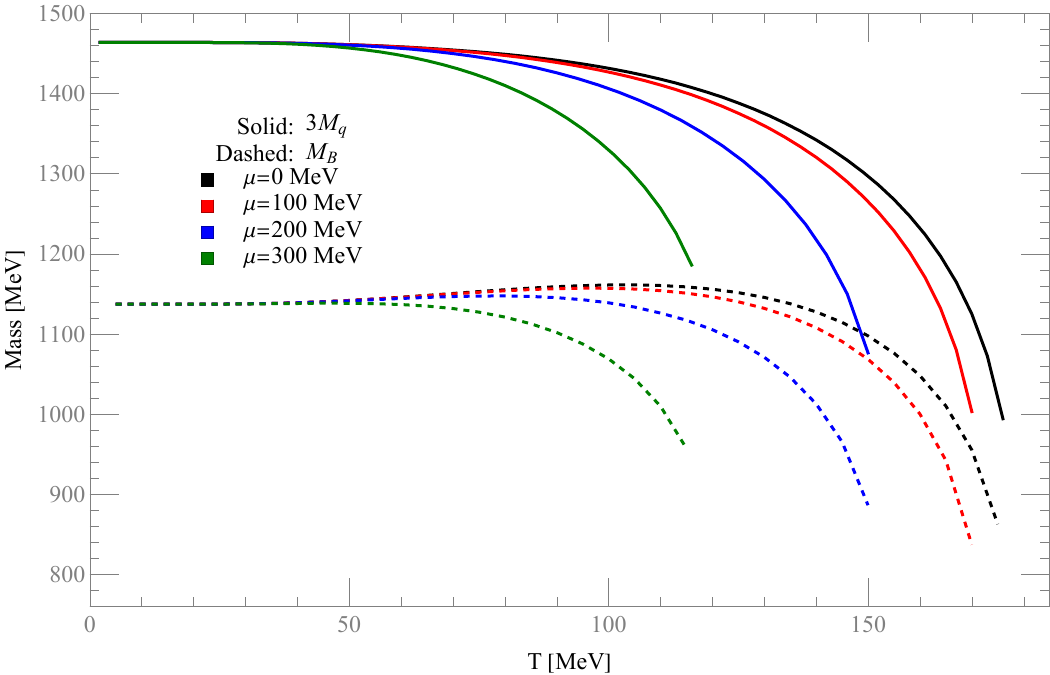}
	\caption{The nucleon mass $M_B$ (dashed lines) and three free constituent quark mass $3M_q$ (solid lines) as a function of $T$, for $\mu = 0,\ 100,\ 200,\ 300\ \rm{MeV}$. } \label{MB3Mq}
\end{figure}

A crucial consideration is the stability of the triple quark system. In our calculations, a baryon is considered as a stable bound state only if its mass is less than the combined mass of its three constituent quarks. In Fig.~\ref{MB3Mq}, we present the baryon mass and the mass of the three constituent quarks as a function of $T$ for different $\mu$ values, $\mu= 0,\ 100,\ 200,\ 300\ \rm{MeV}$. Both quantities decrease with increasing temperature, and the nucleon mass consistently remains lower than the mass of three constituent quarks, indicating stability. As temperature rises, the gap between $M_B$ and $3M_q$ narrows, suggesting the system's increasing instability. This echoes the rising RMS radius, pointing to the system's gradual destabilization.

\begin{figure}[htb]
	\centering
	\includegraphics[width=.45\textwidth]{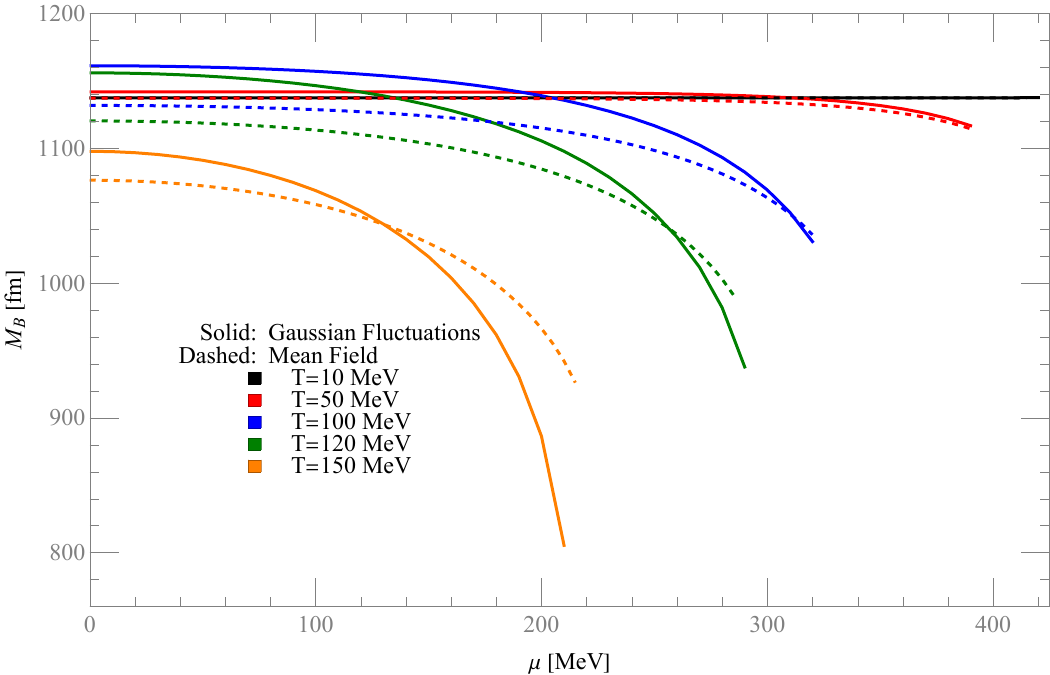}
	\includegraphics[width=.45\textwidth]{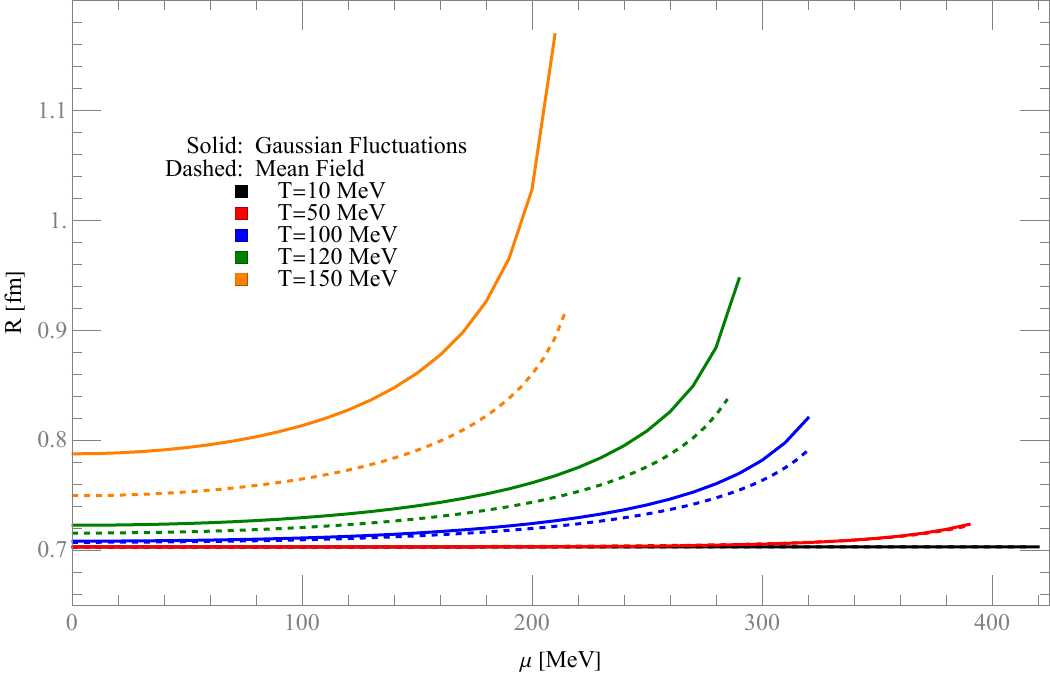}
	\caption{Nucleon mass $M_B$ (left) and RMS radius $R$ (right) as a function of $\mu$  with (solid lines) and without (dashed lines) Gaussian fluctuations, for $T = 10,\ 50,\ 100,\ 120,\ 150\ \rm{MeV}$. } 	
    \label{functionofmu}
\end{figure}

In Fig.~\ref{functionofmu}, we plot the nucleon mass $M_B$ and RMS radius $R$ as a function of chemical potential $\mu$ at different temperatures, $T=10,\ 50,\ 100,\ 120,\ 150\ \rm{MeV}$. Gaussian fluctuations result in a more rapid decrease in $M_B$ and a steeper increase in $R$ compared to the mean-field results, which is similar to the case of finite temperature.


\section{Summary} \label{sec:summary}

In this work, we investigate the quark meson model in a hot and dense background that incorporates Gaussian fluctuations. Following the approach described in Ref.~\cite{Mocsy:2004ab}, meson fluctuations are introduced based on the mean-field theory, and the thermodynamic potential including Gaussian fluctuations is calculated. We derived the equations for the chiral soliton field and the corresponding boundary conditions in this modified background, solving them at various temperatures and densities. The static nucleon properties are calculated by utilizing the chiral soliton solutions.

The presence of Gaussian fluctuations leads to a more rapid expansion of the RMS radius compared to mean-field theory predictions. This effect resembles the Casimir force, known for its repulsive nature, suggesting a comparable phenomenon. Furthermore, the integration of Gaussian fluctuations results in a more rapid decrease in the nucleon mass relative to mean-field results. The non-monotonic trend in the intermediate temperature range arises from a swift rise in the eigenenergy of the quark field. We note that the lightest scalar meson $f_0(500)$ \cite{ParticleDataGroup:2022pth}, which could be regarded as $\sigma$ meson. So we also checked the nucleon properties by decreasing the sigma mass. The nucleon radius always increases due to the existence of Gaussian fluctuations. Since the fluctuation decreases as the sigma mass decreases, the gap of eigenenergy between mean-field and that with fluctuations is smaller and smaller, which results in the non-monotonic behavior of the nucleon mass gradually weakening. The drastic change in hadronic mass at finite temperature and density may play an important role in the hadronization process of heavy-ion collision simulation, which will be left for further research.

\section*{Acknowledgements}
We acknowledge 
the Guangdong Major Project of Basic and Applied Basic Research (Grant No. 2020B0301030008);
the National Natural Science Foundation of China (Grants No. 12105107).

\appendix


\section{Gaussian Fluctuations}

To take the fluctuation field $\Delta,\delta$  average of any complex function  $\mathcal{O}(\sigma_v+\Delta,\delta^2)$, we can use the following steps: Firstly, we expand the function around the point$(\sigma_v,0)$ using Taylor series, and then we take the average term by term.
\begin{eqnarray}
	\langle \mathcal{O}(\sigma_v+\Delta,\delta^2) \rangle =\sum_{k,n} \mathcal{O}^{(k,n)}(\sigma_v,0) \langle \frac{\Delta^k}{k !} \frac{\delta^{2n}}{n !} \rangle,
\end{eqnarray}
where 
\begin{eqnarray}
	\mathcal{O}^{(k,n)}(a,b)=\frac{\partial^k}{\partial a^k} \frac{\partial^n}{\partial b^n} \mathcal{O}(a,b).
\end{eqnarray}

Next, We decompose the vertex $\langle \Delta^k \delta^{2n} \rangle$ into a series of $\left\langle  \Delta^2 \right\rangle $ and $\left\langle  \delta^2 \right\rangle$. We have $\langle \Delta^k \rangle=0$ for odd $k$ and  $\langle \Delta^k \rangle = (k-1)!! \langle \Delta^{2} \rangle^{k/2}$ for even $k$. For triply degenerate pion, we have $\left\langle \delta_1^2\right\rangle = \left\langle \delta_2^2\right\rangle = \left\langle \delta_3^2\right\rangle =\frac{1}{3} \left\langle \delta^2 \right\rangle $, therefore $\left\langle \delta^{2n} \right\rangle = (2n+1)!! \langle \frac{1}{3} \delta^2 \rangle^n $. Finally, replace all the coefficient and sum over all the term. We can find it is equivalent to performing a Gaussian integration on this function as:
\begin{eqnarray}
	\langle \mathcal{O}(\sigma_v+\Delta,\delta^2 ) \rangle = \int_{-\infty}^{\infty} dz P_{\sigma} (z)\int_{0}^{\infty}dy y^2 P_\pi(y) \mathcal{O}(\sigma_v+z,y^2), \label{averge1}
\end{eqnarray}
where
\begin{eqnarray*}
	P_\sigma(z)=\frac{1}{ \sqrt{ 2\pi \langle \Delta^2 \rangle} } \; \mathrm{exp}\left( -\frac{z^2}{2 \langle \Delta^2 \rangle } \right) ,\\
	P_\pi(y) =\sqrt{ \frac{2}{\pi}}\left( \frac{3}{\langle \delta^2 \rangle} \right)^{3/2} \mathrm{exp} \left(\frac{3 y^2}{2 \langle \delta^2 \rangle }\right) .
\end{eqnarray*}

We can also use the same method to expand at point$(\sigma_v,\pi_{1v},\pi_{2v},\pi_{3v})$,
\begin{equation}
	\langle \mathcal{O}(\sigma_v+\Delta,\pi_{1v}+\delta_1,\pi_{2v}+\delta_2,\pi_{3v}+\delta_3 )\rangle = \sum_{k,n} \mathcal{O}^{(k,n,n,n)}(\sigma_v,\pi_{1v},\pi_{2v},\pi_{3v}) \langle \frac{\Delta^k}{k !} \frac{\delta_{1}^{n}}{n !} \frac{\delta_{2}^{n}}{n !} \frac{\delta_{3}^{n}}{n !} \rangle.
\end{equation}
Decompose the fluctuations, we find the form of each part is same. Finally, we get the following equation
\begin{eqnarray}
	\langle \mathcal{O}(\sigma_v+\Delta,\pi_{1v}+\delta_1,\pi_{2v}+\delta_2,\pi_{3v}+\delta_3 ) \rangle&=& \int dz dy_1 dy_2 dy_3 \mathcal{O}(\sigma_v +z,\pi_{1v}+y_1,\pi_{2v}+y_2,\pi_{3v}+y_3) \nonumber \\
	&&P_{\sigma} (z) P_{\pi_1} (y_1) P_{\pi_2} (y_2) P_{\pi_3} (y_3),
\end{eqnarray}
where
\begin{eqnarray*}
	&P_{\pi_1} (y)= P_{\pi_2} (y)= P_{\pi_3} (y)= \frac{1}{ \sqrt{ 2\pi \langle \delta_i^2 \rangle} } \; \mathrm{exp}\left( -\frac{y^2}{2 \langle \delta_i^2 \rangle } \right),  \\
	&\pi^2_{1v}=\pi^2_{2v}=\pi^2_{3v}=\frac{1}{3}\pi_v^2. 
\end{eqnarray*}
Eq.~(\ref{averge1}) holds for any analytic function $\mathcal{O}$, all quantities involving fluctuations must be calculated using this formula. Additionally, we also require the derivative of the function $\left\langle  \mathcal{O}\right\rangle $ with respect to the variable $\mathcal{\alpha}$,
\begin{eqnarray}
	\frac{\partial  }{\partial \alpha}  \langle \mathcal{O}(\sigma_v+\Delta,\delta^2) \rangle &=&\frac{\partial \sigma_v}{\partial \alpha} \langle \frac{\partial \mathcal{O}(\sigma_v+\Delta,\delta^2)}{\partial \sigma_v} \rangle \nonumber \\
	&+&\frac{1}{2} \frac{\partial \langle \Delta^2 \rangle}{\partial \alpha} \langle \frac{\partial \mathcal{O}(\sigma_v+\Delta,\delta^2) }{\partial \Delta^2} \rangle \nonumber\\
	&+&\frac{1}{2} \frac{\partial \langle \delta^2 \rangle }{\partial \alpha} \langle \frac{\partial \mathcal{O}(\sigma_v+\Delta,\delta^2) }{\partial \delta^2} \rangle. 
\end{eqnarray}
Using this identity, we can calculate the meson mass which can be obtained from the second derivative of the thermodynamic potential with respect to fluctuations and chiral condensate, which represents the expectation value of $\sigma$ field. 
\begin{eqnarray}
	m_\sigma^2 &=&\langle \frac{\partial^2 \Omega(\sigma_v +\Delta ,\delta^2 ) }{\partial \Delta^2 } \rangle=\left\langle \frac{\partial^2 U }{\partial \Delta^2} \right\rangle + \left\langle  \frac{\partial^2 \Omega}{\partial \Delta^2} \right\rangle  \nonumber \\
	&=& \lambda (3\sigma_v^2+3\left\langle \Delta^2 \right\rangle +\left\langle \delta^2\right\rangle -\zeta^2 ) + \left\langle  \frac{\partial^2 \Omega}{\partial \Delta^2} \right\rangle, \label{msigma2} \\
	m_\pi^2&=&\langle \frac{\partial^2 \Omega(\sigma_v +\Delta ,\delta^2 ) }{\partial \delta^2 } \rangle=\left\langle \frac{\partial^2 U }{\partial \delta^2} \right\rangle + \left\langle  \frac{\partial^2 \Omega}{\partial \delta^2} \right\rangle  \nonumber \\
	&=& \lambda (\sigma_v^2 +\left\langle \Delta^2\right\rangle +\frac{5}{3}\left\langle \delta^2\right\rangle -\zeta^2 )   + \left\langle  \frac{\partial^2 \Omega}{\partial \delta^2} \right\rangle,  \label{mpi2} \\
	0	&=&\left\langle \frac{ \partial \Omega (\sigma_v +\Delta ,\delta^2 )}{\partial \sigma_v} \right\rangle
	= \left\langle \frac{\partial U }{\partial \sigma_v} \right\rangle + \left\langle  \frac{\partial \Omega}{\partial \sigma_v} \right\rangle  \nonumber \\
	&=& \lambda \sigma_v(\sigma_v^2+3\left\langle \Delta^2\right\rangle+ \left\langle \delta^2\right\rangle -\zeta^2 )-H +\left\langle  \frac{\partial \Omega}{\partial \sigma_v} \right\rangle.  \label{sigmav}  
\end{eqnarray}



\end{document}